\documentstyle[prb,aps,twocolumn]{revtex}
\begin{document}
\draft
\flushbottom
\twocolumn[
\hsize\textwidth\columnwidth\hsize\csname @twocolumnfalse\endcsname

\title{Landau level bosonization of a 2D electron gas}
\author{H. Westfahl Jr.$^1$, A.H. Castro Neto$^2$ and A.O. Caldeira$^1$}
\address{$^{1}$Instituto de F\'{\i}sica Gleb Wataghin\\
Universidade Estadual de Campinas\\
13083-970, Campinas, SP, Brazil\\
$^2$Department of Physics\\
University of California\\
Riverside, CA, 92521, USA}
\date{\today}
\maketitle
\tightenlines
\widetext
\advance\leftskip by 57pt
\advance\rightskip by 57pt

\begin{abstract}
In this work we introduce a bosonization scheme for the low energy
excitations of a 2D interacting electron gas in the presence of an uniform
magnetic field under conditions where a large integral
number of Landau levels are filled.
We give an explicit construction for the electron operator
in terms of the bosons.
We show that the elementary neutral
excitations, known as the {\it magnetic excitons} or {\it magnetoplasma modes},
can be described within a bosonic language and that it provides a
quadratic bosonic Hamiltonian for the interacting electron system
which can be easily diagonalized.
\end{abstract}

\vskip 1cm
\pacs{PACS: 73.40.-c,73.40.Hm, 73.20.Mf, 71.27.+a}

]

\narrowtext
\tightenlines

Bosonization of Fermi fields has been an important non-perturbative tool for
finding the solutions to problems involving many interacting fermions. The
pioneering attempts by Bloch\cite{bloch} on sound waves in dense 1D systems,
which were later extended and given an intuitive foundation by Tomonaga\cite
{tomonaga}, called attention to the conceptual and operational simplicity
that results if all excitations of a fermionic system can be described
within a bosonic language\cite{luttinger,mandelstam,haldane2}. Nowadays the
1D bosonization is a mature field \cite{voit}. Bosonization in higher
dimensions was pioneered by Luther\cite{luther} and extended by Haldane a
few years ago \cite{haldane}. More recently this approach has been applied
in the context of strongly correlated systems where the possibility of
non-conventional ground states seeks for a non-perturbative method of study
\cite{houghton,ah,peter}.

In the context of correlated systems, one of the most interesting phenomenon
in condensed matter physics is the quantum Hall effect \cite
{girvin,kivelson,allan} either in its integer (IQHE) or fractional (FQHE)
regimes. Despite the fact that there is no thermodynamic distinction between
this two phenomena\cite{kivelson}, they have quite different microscopic
origins. The IQHE can be understood from the properties of non-interacting
electrons whereas the understanding of the FQHE necessarily requires a
treatment of the electron-electron interaction. Consequently, almost all the
efforts to study correlations in a 2D electron gas in the presence of a
magnetic field, $B$, were directed toward the study of the strong magnetic
field
case where the FQHE takes place. In this case the Landau level spacing, the
cyclotron energy $\hbar \omega_c = \hbar e B/(mc)$, can be much grater than
the typical Coulomb energy, which is of the order of $e^2/\ell $
(where $\ell =\sqrt{c \hbar/(e B)}$ is the magnetic lenght), and therefore the
system can be described only in terms of the first Landau level and the mixing
with the other Landau levels can be neglected or taken into
account perturbatively. In a weak magnetic field \cite
{allan2} this is not the case and the typical Coulomb energy can exceed the
cyclotron energy resulting in a mixing between the Landau levels. Some
attempts were made to develop a method for studying the mixing caused by
these correlation effects starting from a Fermi Liquid theory \cite
{glazman,levitov} or perturbative Green's function method\cite
{allan2,kallin,fradkin}.

In this paper we start from a Landau level description of the system and
introduce a non-perturbative bosonization scheme for the 2D interacting
fermion gas subject to a perpendicular uniform magnetic field under the
condition that there is a large integer number of filled Landau levels. In this
case, the excitations are electron-hole pairs often called magnetic excitons
and as we will show can be described within a bosonic language.

Recent experiments in  tunneling between parallel 2D electron gases in
applied magnetic fields \cite{eisenstein} have revealed the existence of a
gap in the tunneling density of states. In this case it is established that
correlations have major hole in the tunneling conductance and we expect that
the method presented here can be applied to such problems where the
correlations between the electrons in a magnetic field are so important.

We start with the Hamiltonian of a spinless fermion gas in a plane
perpendicular to an uniform magnetic field
\begin{eqnarray}
H_o=\frac 1{2m}\int d^2{\bf r}\Psi ^{\dagger }({\bf r})\left( {\bf p}-\frac e%
c{\bf A}({\bf r})\right) ^2\Psi ({\bf r}),
\nonumber
\end{eqnarray}
where $\Psi ({\bf r)}$ is the fermion field operator (for the time being we
do not discuss the spin of the electron).

Rewriting $\Psi ({\bf r})$ in a Landau level basis (symmetric gauge) the
Hamiltonian $H_0$ becomes diagonal
\begin{eqnarray}
H_o=\hbar \omega _c\sum_{n,m=0}^\infty (n+1/2)c_{n,m}^{+}c_{n,m}\;\;,
\nonumber
\end{eqnarray}
where $c_{n,m}^{+}$
is the fermion operator that creates a particle in a Landau level $n$ with
the guiding center at a distance $\sqrt{2m+1}\ell $ from the origin of the
coordinate system. Therefore the $m$ degeneracy is related to the location
of the position of the guiding center of the cyclotron orbit. The
non-interacting ground state is generated by uniformly filling $\nu $ Landau
levels ($\nu$ integer) of each guiding center labeled by the quantum number $%
m$
\begin{eqnarray}
\left| G\right\rangle =\prod_{m=0}^{N_\phi }\prod_{n=0}^{\nu
-1}c_{n,m}^{+}\left| 0\right\rangle ,
\nonumber
\end{eqnarray}
where $N_\phi =\frac S{2\pi \ell ^2}$ is the number of flux quanta. The
neutral excitations in the non-interacting system can be generated by
creating a hole in a Landau level $p\leq \nu -1 $ at the guiding center $m$
and an electron in a level $n+p\geq \nu -1$ at the guiding center $m^{\prime }
$.

We note, however, that it is not possible to define eigenstates of the
guiding center $\vec{R}_{0i}$ for the cyclotronic motion of a single charged
particle because of the non-commutativity of the components of the position
operators. Nevertheless we can define a vector $\vec{R}_0=\vec{R}_{0e}+\vec{R%
}_{0h}$, related to the center of mass of the guiding centers of the electron
and the hole excited in the system, whose components $x_0$ and $y_0$
commutes with each other and therefore have simultaneous eigenstates. It is
also possible to define a momentum operator $\vec{P}_0$, canonically
conjugated to $\vec{R}_0$, which is related to the momentum of the center of
mass of the guiding centers. The eigenstates of this operator can be
generated by a superposition of the above described neutral excitations in
different guiding centers and they form a complete basis in the space of the
neutral excitations known as {\it magnetic excitons}\cite{kallin}. In
operator form they can be described as,
\begin{eqnarray}
e_{n,p}^{\dagger }({\bf k})\left| G\right\rangle =e^{-\frac{(k\ell )^2}4%
}\sum_{m^{\prime },m=0}^\infty G_{m^{\prime },m}(\ell k)c_{n+p,m^{\prime
}}^{+}c_{p,m}\left| G\right\rangle ,  \label{m_escit}
\nonumber
\end{eqnarray}
where $k=k_x+ik_y$, $\kappa =|k|$ and $G_{m^{\prime },m}$
is given by\cite{allan}
\begin{eqnarray}
G_{n+p,p}(k)=\sqrt{\frac{p!}{(n+p)!}}\left( \frac{-ik}{\sqrt{2}}\right)
^nL_p^n\left( \frac{\kappa ^2}2\right)
\nonumber
\end{eqnarray}
with $L_p^n$ being generalized Laguerre polynomials.

The density fluctuations of the system can be expressed in terms of
superpositions $a_n^{\dagger }({\bf k})$ of these {\it magnetic excitons}
\begin{eqnarray}
\rho ({\bf k})-\rho _0=\sum_{n=1}^\infty\left\{ a_n^{\dagger }({\bf
k})+a_n(-{\bf k})\right\} ,
\nonumber
\end{eqnarray}
where
\begin{eqnarray}
a_n^{\dagger }({\bf k})=e^{-\frac{(\kappa \ell
)^2}4}\sum_{p=0}^\infty G_{n+p,p}(\ell
k^{*})e_{n,p}^{\dagger }({\bf k})  \label{boson}
\end{eqnarray}
can be interpreted as {\it magnetic plasmons }and $\rho _0=\frac \nu {2\pi
\ell ^2}$ is the electronic average density in the system.

In order to bosonize this problem we proceed exactly like in the Luttinger
model and enlarge the Hilbert space of the fermionic system by including
filled energy eigenstates of negative values of the quantum number $n$. This
construction is the analogue of the sea of infinite negative energy of the
Luttinger model (the Tomonaga construction can also be obtained in an
analogous way by making the Hilbert space finite \cite{paper_longo}). This
vacuum does not concern us because we are studying the {\it low energy
excitation} spectrum of the system. It is clear from this construction that the
bosons only make a faithful representation of the electronic system if $\nu
>>1$. In the same way the bosonization of the electron gas is only possible
when
the system is very dense (large Fermi momentum). By doing that one can show
that the bosonic operators defined in (\ref{boson}) obey the {\it exact}
commutation relations,
\begin{eqnarray}
\left[ a_n({\bf k}),a_{n^{\prime }}^{\dagger }({\bf k}^{\prime })\right]
&=&\delta _{n,n^{\prime }}\delta _{{\bf k,k}^{\prime }}N_\phi {\cal G}%
_n(k\ell )
\nonumber
\\
\left[ a_n({\bf k}),a_{n^{\prime }}({\bf k}^{\prime })\right]  &=&0,
\nonumber
\end{eqnarray}
where
\begin{eqnarray}
{\cal G}_n(x)=nJ_n^2(kR_c), \label{G_n}
\nonumber
\end{eqnarray}
$J_n$ is a Bessel function of the first kind
and $R_c=\sqrt{2\nu} \ell$ is the size
radius of the last occupied ``orbit''. We can proceed further and define the
canonical bosonic operators
\begin{eqnarray}
b_n({\bf k})=\frac 1{\sqrt{N_\phi {\cal G}_n(k\ell )}}a_n({\bf k})
\nonumber
\end{eqnarray}
and recover the standard bosonic commutation relations.

It is easy to show that in terms of these bosonic degrees of freedom the
free Hamiltonian can be rewritten, apart from a background energy, as
\begin{eqnarray}
H_o=\hbar \omega _c\sum_{{\bf k}}\sum_{n=1}^\infty nb_n^{\dagger }({\bf
k})b_n({\bf k})  \label{h0_bos}
\end{eqnarray}
which describes the energy of an assembly of non-interacting bosons.

So far we have succeeded in expressing the neutral excitations of the system
of electrons in terms of bosons. In order to complete our scheme one needs
to express the electron operator in terms of bosons. This is done by
transforming from the momentum representation which is the natural
representation for bosons to the $\left| nm\right\rangle $ representation
which is the appropriate one for fermions in a magnetic field. This should
be contrasted with the case of the electron gas where the momentum is a good
quantum number for the electron and the particle-hole excitation. It means
that we must construct an operator that creates an electron in a Landau
level $n$ and guiding center $m$ as a function of the density fluctuations
within this guiding center. This transformation involves a linear
superposition of the magnetic excitons at different momentum
states. This is done by considering the operator
\begin{eqnarray}
a_{n,m}^{\dagger } &=& \frac
1{\sqrt{n}}\sum_{p=-\infty}^{\infty}c_{n+p,m}^{\dagger }c_{p,m} \nonumber
\\
&=& \int \frac{d^2{\bf k}}{2\pi }e^{-\frac{\kappa ^2}4%
}G_{m,m}(k)b_n^{\dagger }({\bf k}).
\nonumber
\end{eqnarray}
Differently from the operator $b_n^{\dagger }({\bf k})$ that creates density
fluctuations with a definite momentum ${\bf k}$ , the operator $%
a_{n,m}^{\dagger }$ generates density fluctuations on a definite guiding
center. This operator also obeys exact bosonic commutation relations. This
is completely analogous to Luther's \cite{luther,haldane} construction of
the Fermi liquid in 2D and 3D as many 1D fermi liquids, one for each
direction normal to the Fermi surface. But contrary to that construction
here we do not need to ``smear'' the operators close to the Fermi surface.
The discrete nature of the Landau states does this smearing naturally. Here
we have also reduced the 2D problem of electrons in a magnetic field to $%
N_\phi $ 1D {\it chiral} problems, one for each guiding center labeled by $m$.

In order to complete the bosonic description of the fermion operator it will
be necessary to define an operator that creates a fermion in a ``phase''
state associated with this circular motion. This is done by rewriting the
original fermion operator as,
\begin{eqnarray}
c_m(\theta )=\frac 1{\sqrt{2\pi }}\sum_{n=-\infty }^\infty e^{-in\theta}
c_{n,m} \; \; ,
\nonumber
\end{eqnarray}
where we have used the enlarged Hilbert space and it is easy to show that
\begin{eqnarray}
\{c_{m^{\prime }}^{\dagger }(\theta ^{\prime }),c_m(\theta )\} &=&\delta
_{m^{\prime },m}\delta (\theta ^{\prime }-\theta )  \label{anticomutador1} \\
\{c_{m^{\prime }}(\theta ^{\prime }),c_m(\theta )\} &=&0.
\label{anticomutador2}
\end{eqnarray}
The operator $c_m^{\dagger }(\theta )$ creates an electron about the guiding
center given by the quantum number $m$ and at the phase $\theta $ of the
cyclotronic motion. This is the operator that can be properly bosonized.
Indeed the Mandelstam form \cite{mandelstam,haldane2} of the fermion
operator is given by
\begin{eqnarray}
c_m^{\dagger }(\theta )=\frac 1{\sqrt{2\pi }}e^{-i\nu \theta }e^{i\phi
_m^{\dagger }(\theta )}U_me^{i\phi _m(\theta )},
\nonumber
\end{eqnarray}
with
\begin{eqnarray}
\phi _m(\theta )=\frac{N_m\theta }2-i\lim_{\alpha \to 0}\sum_{n=1}^\infty
e^{-\frac{n\alpha }2}\frac{e^{-in\theta }}{\sqrt{n}}a_{n,m}
\nonumber
\end{eqnarray}
and
\begin{eqnarray}
U_m=e^{i\pi \sum\limits_{s=0}^{m-1}N_s}e^{i\Theta _m}
\nonumber
\end{eqnarray}
where the Hermitian operators
\begin{eqnarray}
N_m=\sum\limits_{p=-\infty}^\infty
\left(c_{p,m}^{\dagger}c_{p,m}-<c_{p,m}^{\dagger }c_{p,m}> \right)
\nonumber
\end{eqnarray}
and $\Theta _m$ are canonically
conjugated ($[N_m,\Theta _{m^{\prime }}]=i\delta _{m^{\prime }m}$) such that
the unitary operators $U_m$ anticommute.

In this representation, it can be shown that the fermionic operators obey
the desired anticommutation relations (\ref{anticomutador1}) and (\ref
{anticomutador2}) and it recovers the long time asymptotic behavior of the
non-interacting propagator between two different phase states which is given
by
\begin{eqnarray}
K(\theta ,t) &=&<c_{m^{\prime }}^{\dagger }(\theta ,t)c_m(0,0)>  \nonumber \\
&=&\delta _{m^{\prime },m}\frac{e^{-i\nu \theta }}{1-e^{-i(\theta -\omega
_ct)}}
\nonumber
\end{eqnarray}
and can be related to the space-time propagator by writing the operator $%
\Psi ({\bf r})$ in terms of the $c_m(\theta )$ \cite{paper_longo}. That
completes the construction of the model. Now we have an {\it operator}
relation between electrons in a magnetic field and bosons.

Our next step is to include the interaction between the electrons within
this framework. In an uniform neutralizing background the total Hamiltonian
can be written as
\begin{eqnarray}
H=H_o+H_I
\nonumber
\end{eqnarray}
where $H_o$ is given by (\ref{h0_bos}) and
\begin{eqnarray}
H_I &=&
\frac{1}{\ell ^2} \sum_{{\bf k}} V(k) \sum_{n,n'=1}^\infty \sqrt{%
{\cal G}_n(k\ell ){\cal G}_{n^{\prime }}(k\ell )}  \nonumber \\
&&\left\{ b_n^{\dagger }({\bf k})+b_n(-{\bf k})\right\} \left\{ b_{n^{\prime
}}^{\dagger }(-{\bf k})+b_{n^{\prime }}({\bf k})\right\}
\nonumber
\end{eqnarray}
is a density-density interaction between the fermions which is written in
bosonic language. $V(k)$ is the Fourier transform of the interacting
potential.

In order to diagonalize $H$ we will adopt a generalized Bogoliubov
transformation \cite{ah} with real coefficients $\mu _{nl}({\bf k})$ and $%
\vartheta _{nl}({\bf k})$ which is given by
\begin{eqnarray}
b_n({\bf k}) = \sum_{l=1}^\infty \left( \mu _{nl}({\bf k})\beta _l({\bf
k})+\vartheta _{nl}({\bf k})\beta _l^{\dagger }(-{\bf k})\right) ,
\nonumber
\end{eqnarray}
where $\beta _l({\bf k})$ are the new bosonic operators. This canonical
transformation yields
\begin{eqnarray}
H=\hbar \omega _c\sum_{{\bf k}}\sum_{l=1}^\infty \Omega _l({\bf k})\beta
_l^{\dagger }({\bf k})\beta _l({\bf k})
\nonumber
\end{eqnarray}
where the new eigenfrequencies $\Omega _l({\bf k})$ are determined by
the equation
\begin{eqnarray}
1=\left(\frac{V(k)}{\hbar \omega _c \ell^2}\right)
\sum_{n=1}\frac{{\cal G}_n(k\ell )n}{\left( \Omega _l({\bf k})^2-n^2\right) }%
.  \nonumber
\end{eqnarray}
The above equation has the same form as the one obtained for the normal
modes in the integer quantum Hall effect in the RPA approximation
for a Coulomb potential, $V(\kappa) = 2\pi e^2/\kappa$ \cite
{kallin}. Moreover, the same equation is obtained in the fractional quantum
Hall effect for composite fermions \cite{fradkin}. Now we have shown that as
far as the bosonization construction is valid, the equation which defines
the eigenvalues holds for any value of the magnetic field. This is a
consequence of the fact that bosonization fulfills the Ward identities.
Thus, exactly like in the Fermi gas, the limit of
validity of bosonization goes beyond what is expected.

As a last comment we would like to mention that the method developed in
this letter can be applied to the fractional Hall effect if we use the
Chern-Simons theory to map the original electrons into the composite
fermions. In this case, as it is well established, the fractional quantum
Hall effect is mapped into the integer quantum Hall effect of composite
fermions. The condition that $\nu>>1$ becomes equivalent to say that we are
looking at fillings close to $1/2$ \cite{hlr}. The nature of the ground
state of this system has recently been the source of much controversy and we
believe that our methods can provide non-perturbative answers for the
structure of the ground state of this problem \cite{paper_longo}.

In conclusion, we have proposed a new method of bosonization for 2D fermions
subject to an uniform magnetic field by directly working in a Landau basis.
We propose an analogue of the Luttinger model which can be bosonized and
provides an operator relation between electrons and bosons in the limit of $%
\nu >>1$. This construction is essentially non-perturbative and provides new
insights into the problem of electrons interacting in the presence of
magnetic fields. We show that bosons are collective excitations of the
system which are related with a collection of $N_\phi $ 1D Luttinger liquids
associated with each guiding center. We also obtain the dispersion of the
collective modes of the system and discuss the applicability of the method to
the integer and fractional quantum Hall effects. We expect our method will
enlarge the understanding of the physics of strongly correlated systems in the
presence of magnetic fields.

H.W. and A.O.C. kindly acknowledge respectively the full and partial
supports from the Conselho Nacional de Desenvolvimento Cien\'{i}fico e
Tecnol\'{o}gico (CNPq). We are also grateful to E.H. Fradkin, F.D.M. Haldane
and D. V. Khveshchenko for very useful discussions.

\end{document}